\documentclass[%
 reprint,
superscriptaddress,
groupedaddress,
runinaddress,
showpacs,preprintnumbers,
bibnotes,
 amsmath,amssymb,
 aps,
]{revtex4-1}
\usepackage{graphicx}% Include figure files
\usepackage{dcolumn}% Align table columns on decimal point
\usepackage{bm}% bold math
\usepackage{hyperref}% add hypertext capabilities
\usepackage{setspace}
\usepackage{bbm}

\usepackage[mathlines]{lineno}% Enable numbering of text and display math
\usepackage[dvips]{color}

\begin{document}
\title{Zeeman-field-induced nontrivial topological phases in a one-dimensional spin-orbit-coupled dimerized lattice}% Force line breaks with \\
\author{Masoud Bahari and Mir Vahid Hosseini}
 \email{mv.hosseini@znu.ac.ir}
\affiliation{Department of Physics, Faculty of Science, University of Zanjan, Zanjan 45371-38791, Iran}

\begin{abstract}
We study theoretically the interplay effect of Zeeman field and modulated spin-orbit coupling on topological properties of a one-dimensional dimerized lattice, known as Su-Schrieffer-Heeger model. We find that in the weak (strong) modulated spin-orbit coupling regime, trivial regions or nontrivial ones with two pairs of zero-energy states can be turned into nontrivial regions by applying a uniform (staggered) perpendicular Zeeman field through a topological phase transition. Furthermore, the resulting nontrivial phase hosting a pair of zero-energy boundary states can survive within a certain range of the perpendicular Zeeman field magnitude. Due to the effective time-reversal, particle-hole, chiral, and inversion symmetries, in the presence of either uniform or staggered perpendicular Zeeman field, the topological class of the system is BDI which can be characterized by $\mathbbm{Z}$ index. We also examine the robustness of the nontrivial phase by breaking the underlying symmetries giving rise that inversion symmetry plays an important role.
\end{abstract}


\maketitle
%%%%%%%%%%%%%%%%%%%%%%%%%%%%%%%%%%%%%%%%%%%%%%%%%%%%%%%%%%%%%%%%%%%%%%%%%%%
\section {Introduction} \label{s1}
%%%%%%%%%%%%%%%%%%%%%%%%%%%%%%%%%%%%%%%%%%%%%%%%%%%%%%%%%%%%%%%%%%%%%%%%%%%

Topological phases of matters have attracted a lot of attention in recent years due to discovery of graphene \cite{graphene} and topological insulators \cite{TI}. Also, the search for conventional and unconventional topological superconductors, hosting topological phases \cite{Ts}, has become one of the growing topics in condensed-matter physics \cite{TsExp1,MajorTS}. As long as underlying symmetries in the topological systems are preserved, the symmetry protected gapless edge or surface states within gapped bulk states are robust against many forms of perturbations, making relevant requirement for quantum electronic devices and, especially, topological quantum computations \cite{QuantComput1}.

One-dimensional (1D) heterostructures \cite{1Dhetero1,1Dhetero12,1Dhetero14} consisting of ingredients such as superconductors, topological insulators, spin-orbit coupled semiconductors and ferromagnets can support topologically nontrivial phases hosting, for instance, non-Abelian Majorana bound states at zero energy levels \cite{MajorTS,MajorFermi2} and fractionally charged fermion bound states \cite{divTopo2,fracCharFer,fracCharFer1}. Furthermore, 1D Majorana chain has been proposed to include Majorana Fermions \cite{QuantComput1,kitaevWO1} at its boundaries. In contrast, one of the simplest 1D topological insulators is known as the Su-Schrieffer-Heeger (SSH) model with the BDI symmetry class which is proposed for polyacetylene \cite{SSH1}. SSH model has been shown to exhibit a diversity of nontrivial topological phases \cite{divTopo2,divTopo1}. In this context, most of the recent studies have been devoted to investigating the effects of complex boundary potentials \cite{20}, next-nearest-neighbor hopping \cite{12}, superconducting correlations \cite{13}, disorders \cite{22}, time-dependent potential \cite{FloquetSSH}, magnetic flux \cite{MagFlux}, modulated on-site potentials \cite{18,19}, and hopping amplitudes \cite{19} on topological properties of the 1D topological insulators.

In theoretical and experimental studies, both magnetic field \cite{1Dhetero12,magFiel1} and spin-orbit coupling \cite{spinOrbit1} are important keys for the existence of nontrivial phases. It has been recently shown that a spatially varying magnetic field can be served as an effective spin-orbit interaction leading to creating flat bands of Majorana states \cite{MagToSO1} and fractionalization of charged bound states \cite{fracCharFer1}. It has also been studied the effect of modulated spin-orbit coupling \cite{nonUniformSO} on trivial and nontrivial regimes of SSH model \cite{SoSSH}.
However, the interplay effect of Zeeman fields and spin-orbit coupling on topological properties of 1D superlattices \cite{ZemanSOinter} and, in particular, SSH model has received only limited attention.

In experimentally realizable cases, fine-tuning of parameters of a system to certain values through external fields is not easily reachable, particularly, when dealing with nontrivial topological phases in meso- and nanoscopic platforms \cite{1Dhetero14}. So, it is important to provide a much wider range of required model parameters into reach that facilitates experimental observations of desired phases. Generally speaking, flexibility in increasing or decreasing the range of less important nontrivial or trivial topological phases in favour of a desired topological phase is practically one of the most important requirements of topological quantum computations such that the topological class of the system remains intact.
Therefore, an interesting question is how to increase the nontrivial topological region of 1D dimerized lattice in space of parameters with the combined effects of spin-orbit coupling and Zeeman field respecting underlying symmetries.

In this work, we analyze the effect of perpendicular and parallel Zeeman fields on topological phases of a SSH chain with modulated spin-orbit coupling \cite{SoSSH}. If the Zeeman fields are absent, trivial regions, where the system is an ordinary insulator, can be found in weak and strong modulated spin-orbit coupling regimes. Furthermore, there exist two types of topologically nontrivial phases characterized by one or two pairs of zero-energy edge states. We show that in the presence of uniform (staggered) perpendicular Zeeman field for weak (strong) modulated spin-orbit coupling, the trivial region would be turned into a nontrivial topological one through a topological phase transition by increasing the magnitude of the Zeeman field that results in the emergence of a pair of zero-energy edge states.
Moreover, the perpendicularly applied uniform (staggered) Zeeman field in the weak (strong) modulated spin-orbit coupling regime causes the nontrivial topological phase with two pairs of zero-energy edge states to change its topology and turn into the topologically nontrivial phase hosting one pair of zero-energy states.
Symmetry arguments show that the system in the presence of such perpendicular Zeeman fields possesses inversion, chiral, particle-hole and effective time-reversal symmetries which still falls in the BDI class with topological number $\mathbb{Z}$. We also calculate analytically the winding number through bulk properties of quantum states and show that the results of periodic boundary conditions (PBCs) are in good agreement with those of open boundary conditions (OBCs), according to bulk-edge correspondence. Further, we explore the robustness of the symmetry protected edge states by adding local perturbations such as parallel Zeeman field and spin-dependent on-site potential. These perturbations break the symmetries implying that the nontrivial topological phases are protected by inversion symmetry fundamentally.

This paper is organized as follows: In Sec. \ref{section1}, we introduce a tight-binding model of the system, and obtain bulk energy bands. We also investigate symmetries of the system. In Sec. \ref{section2}, topological phase diagrams without the Zeeman fields are studied. In Secs. \ref{section3} and \ref{section4}, the effects of uniform and staggered perpendicular Zeeman fields, respectively, on the topological phases are studied by calculating of the energy spectrum and topological number of a finite system. For the winding number an analytic formula is derived via bulk states in Sec. \ref{winding}. We discuss the effect of symmetry breaking perturbations on the robustness of edge states in Sec. \ref{section symmetry} and finally conclusions are presented in Sec. \ref{section5}.

%%%%%%%%%%%%%%%%%%%%%%%%%%%%%%%%%%%%%%%%%%%%%%%%%%%%%%%%%%%%%%%%%%%%%%%%%%%
\section {Theoretical Model}\label{section1}
%%%%%%%%%%%%%%%%%%%%%%%%%%%%%%%%%%%%%%%%%%%%%%%%%%%%%%%%%%%%%%%%%%%%%%%%%%%

We consider a 1D dimerized lattice along the $x$-axis which contains two sublattices $a$ and $b$ in each unit cell with a lattice constant $d$. The lattice is subjected to a modulated spin-orbit coupling \cite{SoSSH} and also both perpendicular and parallel Zeeman fields are applied. Thus, the total tight-binding Hamiltonian describing the system is the sum of the Hamiltonians of the SSH model, $\hat{H}_{SSH}$, the spin-orbit coupling, $\hat{H}_{SO}$, and the Zeeman fields, $\hat{H}_{B}$, as
\begin{equation}\label{full hamiltonian}
 \hat{H} = \hat{H}_{SSH}+\hat{H}_{SO}+\hat{H}_B,\\
\end{equation}
with
\begin{eqnarray}\label{hamiltonian-real space}
\hat{H}_{SSH}\!&=&\!-\!\!\sum\limits_{n,\sigma}[(t-\delta t)\hat{a}^\dagger_{n,\sigma}\hat{b}_{n,\sigma}
\!+\!(t+\delta t)\hat{a}^\dagger_{n,\sigma}\hat{b}_{n-1,\sigma}\!+\!h.c.],\nonumber \\
\hat{H}_{SO}&=&\sum\limits_{n,\sigma}[\lambda\hat{a}^\dagger_{n,\sigma}\hat{b}_{n,-\sigma}-\lambda^\prime\hat{a}^\dagger_{n,\sigma}\hat{b}_{n-1,-\sigma}+h.c.],\nonumber\\
\hat{H}_B&=&\sum_{n,\sigma,\sigma^\prime}[\hat{a}^\dagger_{n,\sigma}(\mathbf M_{n,a}\cdot\mathbf{\tau})_{\sigma\sigma^\prime}\hat{a}_{n,\sigma^\prime}\nonumber\\ &+&\hat{b}^\dagger_{n,\sigma}(\mathbf M_{n,b}\cdot {\bf\tau})_{\sigma\sigma^\prime}\hat{b}_{n,\sigma^\prime}],\nonumber
\end{eqnarray}
where $\hat{a}^\dagger_{n,\sigma} (\hat{a}_{n,\sigma})$ and $\hat{b}^\dagger_{n,\sigma}$ ($\hat{b}_{n,\sigma}$) are the fermion creation (annihilation) operators of electrons with spin $\sigma = (\uparrow \textrm{or} \downarrow)$ on the sublattices $a$ and $b$ of the {\it n}th unit cell, respectively. In addition, $t - \delta t$ and $t + \delta t$ ($\lambda$ and $\lambda^\prime$) denote the hopping (spin-orbit coupling) amplitudes in the unit cell and between two adjacent unit cells, respectively, with the dimerization strength $\delta t$, $\mathbf\tau$ is the Pauli vector acting on the spin subspace and the Zeeman field vector on the {\it n}th unit cell is $\mathbf M_{n,i} = (M_{n,x,i}, M_{n,y,i}, M_{n,z,i})$ with sublattice index $i =( a\quad \textrm{or} \quad b )$. Here, since the lattice is invariant under rotations about the $x$-axis, without loss of generality, we apply the perpendicular Zeeman field along the $y$-direction, {\it i.e.}, $M_{n,z,i} = 0$. Furthermore, for simplicity, the Zeeman field is supposed to be identical in each unit cell, so we drop its unit cell index $n$ hereafter. We shall later investigate the effect of the Zeeman field in the $x$-direction, therefore we include the Zeeman field as $\mathbf M_{i} = (0, M_{y,i}, 0)$. We choose $t$ as the energy unit and the lattice constant $d = 1$.
%
%%%%%%%%%%%%%%%%%%%
\begin{figure}
 \centering
  \includegraphics[width=8.5cm]{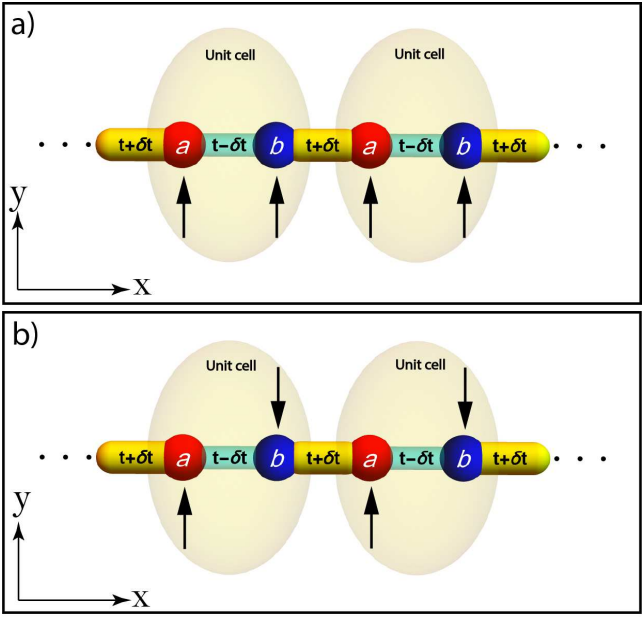}
  \caption{(Color online) Schematic illustration of a 1D dimerized lattice consisting of two sublattices (blue and red balls) with perpendicularly applied Zeeman field (black arrows). Configuration of (a) uniform perpendicular Zeeman field for weak modulated spin-orbit coupling regime and (b) staggered perpendicular Zeeman field for strong modulated spin-orbit coupling regime. $t - \delta t$ $(t + \delta t)$ indicates inter-cell (intra-cell) hopping energy.}
  \label{fig1fig2}
\end{figure}

Adopting PBCs and Fourier transforming, the total Hamiltonian $\hat{H}$, Eq. (\ref{full hamiltonian}), can be easily written as
\begin{equation}\label{hk}
\hat{H}=\sum_{k}\hat{\psi}_k^{\dagger}\hat{h}(k)\hat{\psi}_k,
\end{equation}
where $\hat{\psi_k^\dagger} = (\hat{a}_{k,\uparrow},\hat{a}_{k,\downarrow},\hat{b}_{k,\uparrow},\hat{b}_{k,\downarrow})^\dagger$, and
\begin{equation}\label{hk-matrice}
\hat{h}(k)=\left(
\begin{array}{cccc}
     0&-iM_{y,a}&s(k)&\xi(k)\\
     iM_{y,a}&0&\xi(k)&s(k)\\
     s(k)^\ast&\xi(k)^\ast&0&-iM_{y,b}\\
     \xi(k)^\ast&s(k)^\ast&iM_{y,b}&0\\
   \end{array}
   \right),
\end{equation}
with
\begin{align}\label{hopping and spin-orbit relation}
s(k) = -(t-\delta t)-(t+\delta t) e^{-ik}, \quad &\xi(k)=\lambda-\lambda^\prime e^{-ik}.\\ \nonumber
\end{align}
We diagonalize the Hamiltonian in the momentum space, Eq. (\ref{hk-matrice}), to obtain the eigenvalues as
\begin{equation}\label{energy espetrum}
E(k) = l\sqrt{\frac{\alpha(k)+m\sqrt{\beta(k)}}{2}},
\end{equation}
with
\begin{eqnarray}
\alpha(k)&=& M_{y,a}^2+M_{y,b}^2+2ss^\ast+2\xi \xi^\ast,\nonumber\\
\beta(k)&=& M_{y,a}^4+M_{y,b}^4+M_{y,a}M_{y,b}(8(ss^\ast-\xi \xi^\ast)-2)\nonumber\\
&+&(M_{y,a}+4M_{y,b})(ss^\ast+\xi \xi^\ast)+4(s^\ast\xi+s\xi^\ast)^2,\nonumber
\end{eqnarray}
where $l = + (-)$ indicates the conduction (valance) band and $m = + (-)$ represents upper (lower) subband.

It is well studied that a topological phase transition accompanying by closing and reopening of bulk band gap always determines the boundaries between different topological phases in the absence of interactions \cite{NoGapClosing}. The gap closing conditions of $\hat{h}(k)$ can be obtained by calculating $\det(\hat{h}(k)) = 0$. In the absence of both the perpendicular Zeeman field and the modulated spin-orbit coupling, the energy spectrum reduces to $E(k) = \pm\sqrt{ss^\ast}$ and the energy gap closes at the momentum $k = \pi$ if $\delta t = 0$. On the other hand, in the presence of the modulated spin-orbit coupling and $\mathbf M_{i} = (0, 0, 0)$, the energy gap closure conditions are \cite{SoSSH}
\begin{equation}
2|\delta t| = |\lambda + \lambda^\prime|,
\label{gap clo k=0}
\end{equation}
and
\begin{equation}
2|t|=|\lambda - \lambda^\prime|,
\label{gap clo k=pi}
\end{equation}
at the boundaries of the Brillouin zone $k = \pm\pi$ and momentum $k = 0$, respectively. But in the presence of both the $y$-component of Zeeman field and the modulated spin-orbit coupling, we find that the gap closes at the momenta $k = \pm\pi$  and $k = 0$ with the corresponding conditions, respectively, given by
\begin{equation}
M_{y,a}M_{y,b} = 4\delta t^2 - (\lambda + \lambda^\prime)^2,
\label{gap closure k=0}
\end{equation}
and
\begin{equation}
M_{y,a}M_{y,b} = 4t^2 - (\lambda - \lambda^\prime)^2.
\label{gap closure k=pi}
\end{equation}
In fact, the above relations are the generalized versions of Eqs. (\ref{gap clo k=0}) and (\ref{gap clo k=pi}). This means that the boundaries between topological phases can be changed drastically, as a result of the perpendicular Zeeman field.
Importantly, the sign of right hand sides of Eqs. (\ref{gap closure k=0}) and (\ref{gap closure k=pi}) can be either positive, namely,
\begin{align}
4\delta t^2 > (\lambda + \lambda^\prime)^2,\ \textrm{and}\ 4t^2 > (\lambda - \lambda^\prime)^2,
\label{Weakspin-orbit}
\end{align}
or negative, namely,
\begin{align}
4\delta t^2& < (\lambda + \lambda^\prime)^2,\ \textrm{and}\ 4t^2 < (\lambda - \lambda^\prime)^2,
\label{Strongspin-orbit}
\end{align}
depending on the values of the parameters. We refer to the former (latter) case, {\it i.e.}, relation (\ref{Weakspin-orbit}) ((\ref{Strongspin-orbit})) as the weak (strong) modulated spin-orbit coupling case. Consequently, the weak (strong) modulated spin-orbit coupling requires $M_{y,a}M_{y,b} > 0$ ($M_{y,a}M_{y,b} < 0$). In addition, there are some parameter regions where only one of the inequalities from each relations (\ref{Weakspin-orbit}) and (\ref{Strongspin-orbit}) can be fulfilled. We use the phrase `mixed regime' to refer to these cases. Therefore, the perpendicular Zeeman fields of the sublattices $a$ and $b$ should be either parallel or antiparallel with respect to each other in the unit cell as depicted in Fig. \ref{fig1fig2}. Subsequently, one may anticipate that uniform and staggered Zeeman fields not only satisfy the energy gap closing conditions, [see Eqs.(\ref{gap closure k=0}) and (\ref{gap closure k=pi})] but also preserve some symmetries of the system which will be discussed below.

The system shows particle-hole and chiral symmetries defined respectively as $\mathcal{P}\hat{h}(k)\mathcal{P}^{-1} = -\hat{h}(-k)$ and $\mathcal{C}\hat{h}(k)\mathcal{C}^{-1} = -\hat{h}(k)$  where $\mathcal{P} = \sigma_z\otimes I \mathcal{K}$ and $\mathcal{C} = \sigma_z\otimes\tau_x$ with $I$, $\mathcal{K}$, and $\tau_x$ being the identity matrix, the complex conjugation, and the $x$-component of the Pauli vector $\mathbf{\tau}$, respectively. Also, $\sigma_z$ is the $z$-component of the Pauli vector $\mathbf{\sigma}$ acting on sublattice space. Because of the fact that $\mathcal{T}\cdot \mathcal{P}  = \mathcal{C}$, the anti-unitary effective time-reversal operator can be determined as $\mathcal{T} = I\otimes\tau_x \mathcal{K}$. These symmetry operators also exhibit the features that $\mathcal{P}^2 = 1$, $\mathcal{C}^2 = 1$, and $\mathcal{T}^2 = 1$. As a result of the last property, there is no Kramer's degeneracy related to effective time-reversal symmetry. Therefore, according to the standard Altland-Zirnbauer classification \cite{Altland}, the symmetry class is BDI in the presence of either uniform, $M_{a,y} = M_{b,y}$, or staggered, $M_{a,y} = -M_{b,y}$, Zeeman field. Moreover, when $M_{a,y} = M_{b,y}$, the Hamiltonian (\ref{hk-matrice}) also satisfies the inversion symmetry relation $\mathcal{I}_{u}\hat{h}(k)\mathcal{I}_{u}^{-1} = \hat{h}(-k)$, where $\mathcal{I}_{u} = \sigma_x\otimes I$ with $\sigma_x$ being the $x$-component of $\mathbf{\sigma}$. On the other hand, if $M_{a,y} = - M_{b,y}$, it can be clearly seen that $\hat{h}(k)$ fulfils $\mathcal{I}_{s}\hat{h}(k)\mathcal{I}_{s}^{-1} = \hat{h}(-k)$ under the inversion symmetry operation $\mathcal{I}_{s} = \sigma_x\otimes\tau_x$. Notice that due to inversion and reflection symmetries, topological classifications fall beyond the standard classification \cite{Altland} which are based on the global symmetries, {\it i.e.}, time-reversal, particle-hole, and sublattice (chiral) symmetries. In cases satisfying to the modified topological classifications \cite{modifiedClass,30}, the inversion symmetry operator commutes with all the global symmetries \cite{30}. In contrast, in our case, interestingly, the inversion operators $\mathcal{I}_{u}$ and $\mathcal{I}_{s}$  anti-commute with both chiral and particle-hole operators but commute with effective time-reversal operator. Note that a one-dimensional system belonging to the BDI class has either $\mathbb{Z}$ or $\mathbb{Z}_2$ topological index depending on algebraic relations between inversion operator and other global symmetry operators \cite{modifiedClass,ModiAlgebra}. According to the modified periodic table \cite{modifiedClass,ModiAlgebra}, the algebraic relations of our system imply that topological index is $\mathbb{Z}$.
\section{Topological phases of the 1D chain without perpendicular Zeeman field}\label{section2}
In order to understand the effect of the perpendicular Zeeman fields on the topological properties of the system, the topological phase diagram in the absence of such fields is first investigated by calculating topological invariant \cite{TopoInvar1,TopoInvar3} numerically. The relevant topological invariant for 1D system can be expressed by the topological number $\mathbbm{Z} = \phi_{Zak}/\pi$ where the Zak phase \cite{Zak,40} is $\phi_{Zak} = \sum_{E<0}\int_{-\pi}^{\pi} \langle u_k|i\partial_k u_k\rangle\,dk$ with $|u_k\rangle$ being the occupied Bloch states \cite{TopoInvar3}. Both quantized Zak phase \cite{quanZak1,quanZak2} and symmetry-protected states localized on the boundaries of the system \cite{zero-edg} characterize the existence of nontrivial topological phase, emerging from bulk states properties and symmetry configurations. The quantized Zak phase has also been measured experimentally in 1D periodic potentials using ultracold atoms in optical lattices \cite{ZakExperCold} and periodic tubes in phononic crystals \cite{ZakExperAcou}.

\begin{figure}[h]
  \centering
  \includegraphics[width=8.5cm]{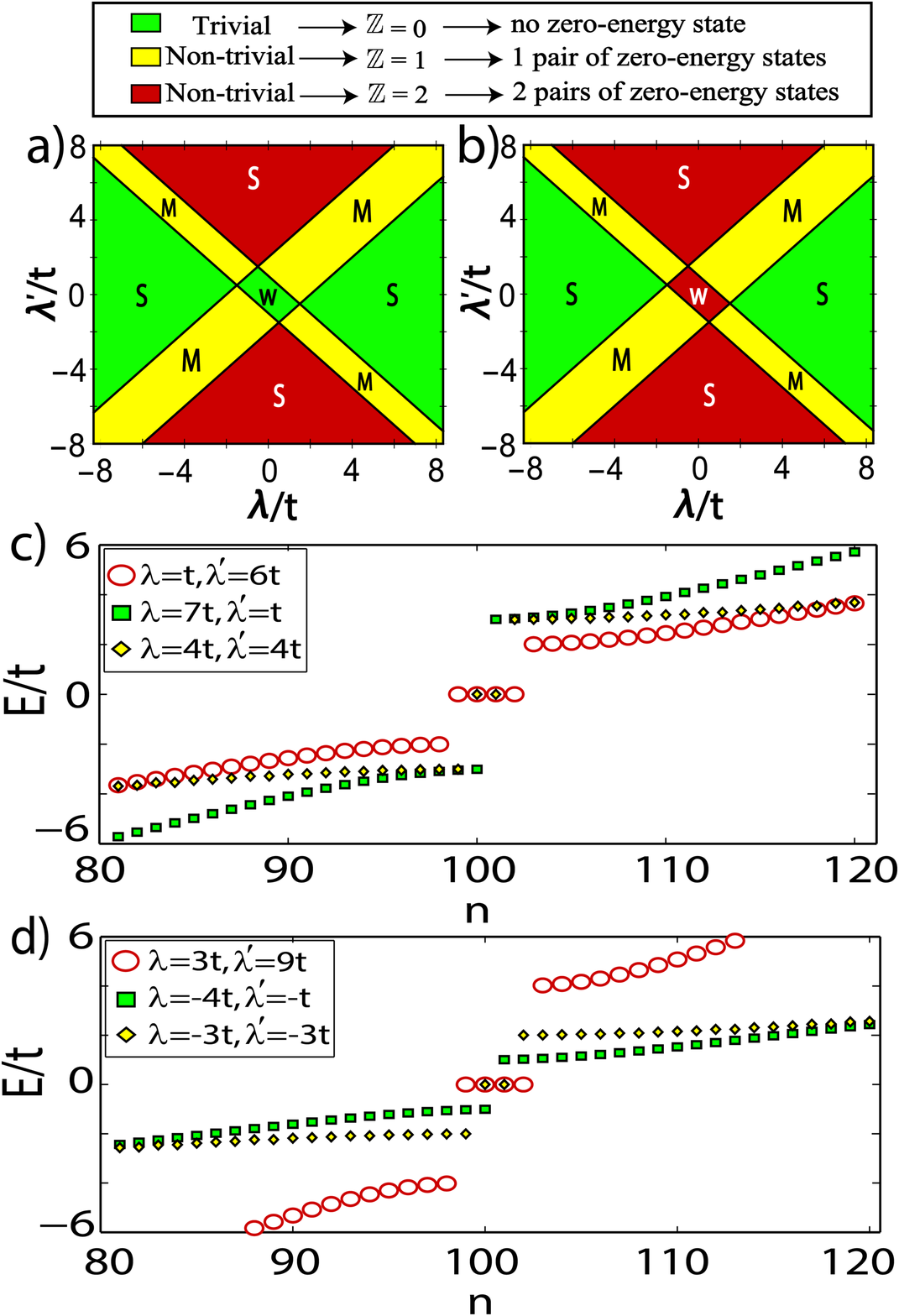}
    \caption{(Color online) Phase diagram in the plane $(\lambda, \lambda^\prime)$ without Zeeman fields for parameters: (a) $\delta t = -0.5t$, (b) $\delta t = 0.5t$. Capital letters M, S, and W denote mixed, strong, and weak modulated spin-orbit coupling regimes. Topologically distinct regions $\mathbbm{Z} = 0, 1,$ and $2$ are depicted by gray (green), light (yellow), and dark (red) colors, respectively. The relevant part of energy spectrum for different values of spin-orbit couplings under OBCs with 50 unit cells is shown for (c) $\delta t = -0.5t$ and (d) $\delta t = 0.5t$. In the middle of the energy gap, the spectrum contain zero-, two-, or four-fold degenerate states corresponding to topologically distinct insulating phases.}
    \label{fig2}
\end{figure}

Since the value of Zak phase is gauge dependent, we follow the choice of unit cell so that the Zak phase of Bloch bands takes the values $0$, $\pi$ or $2\pi$ \cite{quanZak1,gaugeZak}. Thus, the topological number $\mathbbm{Z} = 1 (2)$ corresponds to the existence of one pair (two pairs) of zero-energy states at the boundaries under OBCs and indicates that the system is in the topologically nontrivial phase. Furthermore, when the number $\mathbbm{Z} = 0$, the system is topologically trivial without zero-energy states.

\begin{figure*}[t]
  \centering
  \includegraphics[width=.7\linewidth]{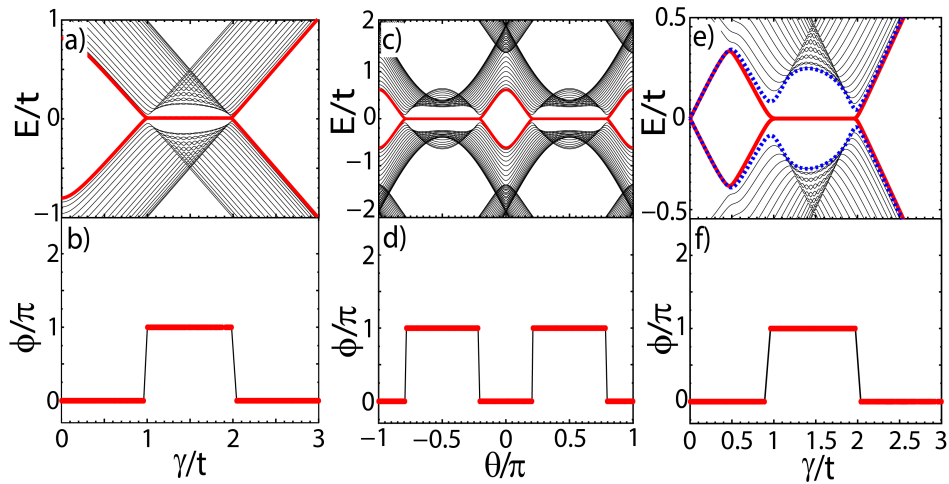}
    \caption{(Color online) Top panels: Dependence of energy spectrum of finite lattice with 20 unit cells on (a) ((e)) $\gamma$ with $\delta t = -(+)0.5t$, $\lambda = \lambda^\prime = 0.1 (0.2)t$ and $\theta = \pi/2$ and on (c) $\theta$ with $\delta t = -0.5t$, $\lambda = \lambda^\prime = 0.1t$ and $\gamma=1.6t$. Bottom panels (b), (d), and (f) are the corresponding Zak phases of the top panels. Here, $\phi_a = 0$.}
    \label{fig3}
\end{figure*}

The resulting topological phase diagrams in the plane $(\lambda, \lambda^\prime)$ for the case $\delta t < 0$ and $\delta t > 0$ are shown in Figs. \ref{fig2}(a) and \ref{fig2}(b), respectively. The black solid lines are phase boundaries
where the bulk band gap closes. Topologically distinct regions $\mathbb{Z} = 0, 1,$ and $2$ are indicated by gray (green), light (yellow), and dark (red) colors, respectively. The different regimes of mixed, weak, and strong modulated spin-orbit couplings are denoted by ``$M$", ``$W$", and ``$S$", respectively, as well. As one can see in Fig. \ref{fig2}(a), there are three trivial regions, $\mathbb{Z} = 0$. While one of them is located in the weak modulated spin-orbit regime, the others are in the the strong modulated spin-orbit regime. Also, in the strong modulated spin-orbit and mixed regimes, there are nontrivial topological phases with different topological number $\mathbb{Z}$. For the case of $\delta t > 0$, there are two trivial regions in the strong modulated spin-orbit regime and the nontrivial topological phase with topological number $\mathbb{Z} = 2$ can be found not only in the strong modulated spin-orbit regime but also in the weak modulated spin-orbit regime [see Fig. \ref{fig2}(b)]. In both cases $\delta t < 0$ and $\delta t > 0$, the topological phase with number $\mathbb{Z} = 1$ is only restricted to the mixed regime. Notice that if $\delta t \neq 0$, the phase diagrams are not symmetric with respect to the parameters $\lambda$ and $\lambda^{\prime}$. To show the number of zero-energy states in connection with the distinct topological phases, we calculate the energy spectra of the 1D lattice dimerization with 50 unit cells under OBCs for the cases $\delta t < 0$ and $\delta t > 0$ as presented in Figs. \ref{fig2}(c) and \ref{fig2}(d), respectively. In either cases, in the middle of the energy gap, depending on the appropriate choice of parameters $\lambda$ and $\lambda^{\prime}$ in the three different regions of panels (a) and (b), one can observe zero, one, or two pairs of zero-energy states related to topologically distinct insulating phases.

\section{The effect of uniform perpendicular Zeeman field on the topological phases}\label{section3}

We consider the $y$-components of Zeeman field terms on sublattices $a$ and $b$ which vary cyclically with the parameter $\theta$ as,
\begin{equation}
  M_{y,a} = \gamma_a \sin(\theta + \phi_a), \quad M_{y,b} = \gamma_b \sin(\theta + \phi_b),
  \label{zeeman fields}
\end{equation}
where the amplitude $\gamma_{a(b)}$ and phase factor $\phi_{a(b)}$ are parameters to control the strength and sign of the Zeeman field on sublattice $a$ ($b$), respectively. Without loss of generality, we assume $\phi_{b} = 0$. Also, we take $\gamma_a = \gamma_b = \gamma$ and $\phi_a=2n^\prime\pi$ or $\phi_a=(2n^\prime-1)\pi$ with $n^\prime$ an integer, as required by the inversion symmetry.

As discussed already above, when $\delta t < 0$, in the absence of Zeeman fields, the system is always gapped and subsequently, a trivial insulator in the weak modulated spin-orbit coupling regime. Thus, in order to close the energy gap and change the topology of the band structure, we must apply a uniform perpendicular Zeeman field [see Eqs. (\ref{gap closure k=0}) and (\ref{gap closure k=pi})]. It's straightforward to see that for $\phi_a = 2n^\prime\pi$, which establishes the uniform Zeeman field, closing and reopening of the bulk gap take place at two certain values of the Zeeman field magnitude given by
\begin{eqnarray}
\gamma^{u}_{1}&=&\sqrt{4\delta t^2-(\lambda+\lambda^\prime)^2},\label{uniZeeman1}\\
\gamma^{u}_{2}&=&\sqrt{4t^2-(\lambda-\lambda^\prime)^2},\label{uniZeeman2}
\end{eqnarray}
within which this uniform Zeeman field drives the system into topologically nontrivial phase.
Therefore, the topological phases caused by Zeeman field can be characterized by the $\mathbbm{Z}$ such that the quantized Zak phase $\pi$ ($\mathbbm{Z} = 1$) manifests the existence of a pair of zero-energy edge states under OBCs.

The dependence of energy spectrum on $\gamma$ is shown in Fig. \ref{fig3}(a) for $\delta t < 0$ with $\theta = \pi/2$ and $\lambda = \lambda^\prime = 0.1t$, {\it i.e.}, weak modulated spin-orbit coupling regime. Low energy states associated with edge states are indicated by thick solid (red) lines. One can see that within a certain range of $\gamma$ the lowest energy states become zero-energy states. In this region, the value of Zak phase is $\pi$ as shown in Fig. \ref{fig3}(b) indicating the establishment of nontrivial topological phase. Outside of this region, the Zak phase takes $0$ value and the energy gap is open as well. The induced nontrivial topological phase, due to perpendicular Zeeman field, stems from spin-splitting of the helical basis projected into a 1D bipartite lattice.

\begin{figure}[b]
  \centering
  \includegraphics[width=1\linewidth]{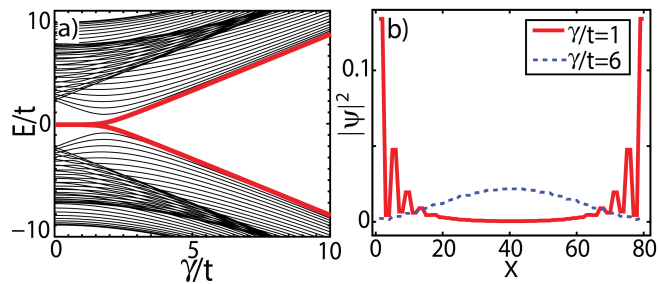}
    \caption{(Color online) (a) The energy spectrum as a function of $\gamma$ under OBCs in presence of uniform Zeeman field with 20 unit cells. (b) Probability distribution for two different values $\gamma/t = 1$ and $6$. Here, $\delta t = -0.5t$, $\theta = \pi/2$, $\lambda = \lambda^\prime = 4t$, and $\phi_a = 0$.}
    \label{fig4}
\end{figure}

The energy spectrum as a function of $\theta$ is plotted in Fig. \ref{fig3}(c) for $\delta t < 0$ with parameters $\gamma = 1.6t$ and $\lambda = \lambda^\prime = 0.1t$. There is an energy gap for small $\theta$. As we increase $\theta$ in magnitude, a phase transition occurs, consequently, the gap closes and flat bands appear. With further increase of $\theta$, flat bands disappear and gap reopens. Naturally, the spectrum is symmetric about $\theta = 0$ arising from reflection symmetry with respect to the $xz$-plane. The corresponding Zak phase is shown in Fig. \ref{fig3}(d). One can see that in the parameter region in which the flat bands are appeared, the Zak phase takes $\pi$ value indicating the existence of zero-energy edge states, whereas, in the other region, the Zak phase is 0 exhibiting a topologically trivial phase.

\begin{figure*}
  \centering
  \includegraphics[width=.7\linewidth]{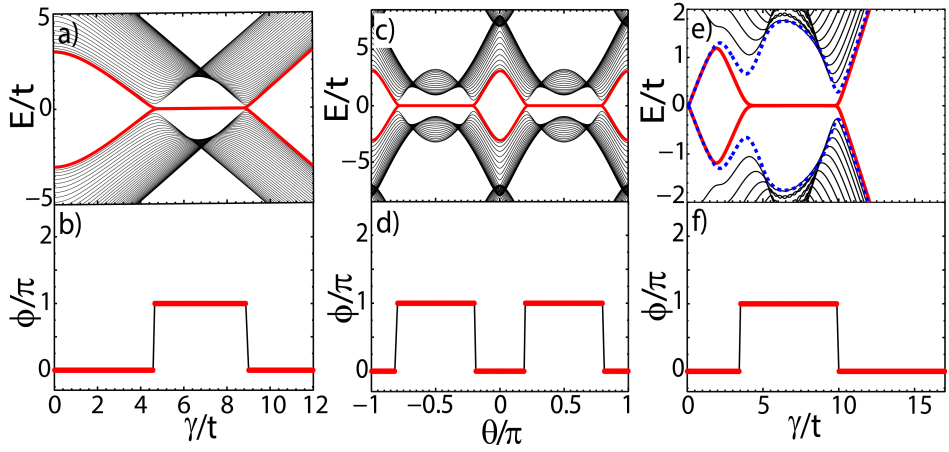}
    \caption{(Color online) Top panels: Dependence of energy spectrum of a finite lattice with 30 unit cells on (a) ((e)) $\gamma$ with $\delta t = +(-)0.5t$, $\lambda = 7 (3)t$, $\lambda^\prime = 2 (7)t$ and $\theta = \pi/2$ and on (c) $\theta$ with $\delta t = 0.5t$, $\lambda = 7t$, $\lambda^\prime = 2t$ and $\gamma = 8t$. Bottom panels (b), (d), and (f) are the corresponding Zak phases of the top panels. Here, $\phi_a = \pi$.}
    \label{fig5}
\end{figure*}

Upon applying the uniform perpendicular Zeeman field, both pairs of edge states for the case of $\delta t > 0$ in the weak spin-orbit coupling become unstable and gap opens, as shown in Fig. \ref{fig3}(e). With the increase of the $\gamma$, while one of the pair states remains unstable, interestingly, the other one becomes stable; as a result, the system undergoes a topological phase transition from a normal insulator to a topological insulator. Moreover, at large $\gamma$, since only one of the spin species will be dominated, the possibility of establishing the edge states containing both types of spins states vanishes and subsequently gap opens for the second time. The corresponding Zak phase as a function $\gamma$ is illustrated in Fig. \ref{fig3}(f). One can see that if the zero-energy states are present, then the Zak phase shows a nontrivial topology.
It is worthwhile noting that, in the case of weak spin-orbit coupling for both $\delta t < 0$ and $\delta t > 0$, if $\phi_a = (2n^\prime - 1)\pi$, one obtains $\phi_{Zak} = 0$ ($\mathbbm{Z} = 0$), thus, the system is always gapped.

In Fig. \ref{fig4}(a), the energy spectrum versus $\gamma$ is shown for mixed regime with parameters $\delta t =-0.5t$, $\lambda = \lambda^\prime = 4t$, and $\theta = \pi/2$ (the result of case $\delta t  >  0$ is the same and not shown). In this regime, for both cases $\delta t > 0$ and $\delta t < 0$, the uniform perpendicular Zeeman field preserves the two-fold degenerate zero-energy edge state from zero up to a certain value of the Zeeman field strength. This can be understood as follows. In the mixed region, equal strengths of inter- and intra cell spin-orbit couplings with almost large magnitudes cause that the band structure around the Fermi surface can be affected slightly by small or mediate Zeeman field strengths without losing its topological features. As described above, increasing the Zeeman strength beyond the critical value, completely splits opposite spin states resulting in the disappearance of zero-energy states. In order to confirm that the zero-energy states exhibit themselves as localized states at sample boundaries, the probability distribution of the lowest energy states for two different values $\gamma = t$ and $6t$ with the same parameters as Fig. \ref{fig4}(a) is depicted in panel (b). In the presence of zero-energy states, for $\gamma = t$, the probability distribution has maxima at the two boundaries of the chain, while for $\gamma = 6t$, minimum values of probability distribution occur at such points.

\section{The effect of staggered perpendicular Zeeman field on the topological phases}\label{section4}

In this section we will first investigate the effect of perpendicular Zeeman field on different topological phases occurred in the strong modulated spin-orbit coupling regime. As discussed in Sec. \ref{section2}, in the absence of Zeeman fields, for both cases $\delta t < 0$ and $\delta t > 0$, when inter-cell spin-orbit coupling is larger than intra-cell one the system is a trivial insulator. To change the topology of {these} regions from trivial phase into a nontrivial one, it is necessary to close and then reopen the energy gap with the help of staggered perpendicular Zeeman field [see Eqs. (\ref{gap closure k=0}) and (\ref{gap closure k=pi})]. It can be easily checked that for $\phi_a = (2n^\prime - 1)\pi$, providing the staggered Zeeman field, within the range $\gamma^{s}_{1}< M_{y,a} < \gamma^s_{2}$ the phase of trivial region turns into the nontrivial. Here, we have defined,
\begin{eqnarray}
\gamma^{s}_{1}&=& \sqrt{(\lambda - \lambda^\prime)^2 - 4t^2},\label{stagZeeman1}\\
\gamma^{s}_{2}&=& \sqrt{(\lambda + \lambda^\prime)^2 - 4\delta t^2}\label{stagZeeman2}.
\end{eqnarray}

The evolution of energy spectrum as functions of $\gamma$ and $\theta$ is depicted in Figs. \ref{fig5}(a) with $\theta = \pi/2$  and \ref{fig5}(c) with $\gamma = 8t$, respectively, for $\delta t = 0.5t$, $\lambda = 7t$, $\lambda^\prime = 2t$, and $\phi_a = \pi$. The bulk states of the eigenvalues, depicted by thin solid black lines, have a different structure compared to the uniform perpendicular Zeeman field case [see Figs. \ref{fig3}(a) and \ref{fig3}(c)], whereas the lowest energy states of eigenvalues (thick solid red lines) including the boundary states reveal a broadly similar behavior. Also, the evolution of Zak phase as functions of $\gamma$ and $\theta$ with the same parameters as Figs. \ref{fig5}(a) and \ref{fig5}(c) is presented in Figs. \ref{fig5}(b) and \ref{fig5}(d), respectively. They show that zero-energy states have nontrivial characteristics.
Also, the $\gamma$ dependence of energy spectrum in the strong modulated spin-orbit coupling regime containing four-fold degenerate boundary states is shown in Fig. \ref{fig5}(e). As can be seen, there is only a four-fold degenerate zero-energy state at $\gamma = 0$. Similar to Fig. \ref{fig3}(e), as we increase the strength of staggered perpendicular Zeeman field, the system goes to a trivial phase such that the lower and upper bulk bands separate with an energy gap in the parameter regime $\gamma \in(0,3.46t)$. By further increasing the strength of Zeeman field, two branches of low energy states (thick solid red lines) merge together again in the parameter regime $\gamma \in(3.46t,9.94t)$ as zero-energy flat bands ($\mathbb{Z} = 1)$. Thus, the unstable four-fold degenerate nontrivial topological phase in the strong modulated spin-orbit coupling regime for $\delta t < 0$ (even for $\delta t > 0$) is changed to the two-fold degenerate topologically nontrivial phase by $\gamma$ though undergoing a topological phase transition. Also, in Fig. \ref{fig5}(f), the $\gamma$ dependence of Zak phase of panel (e) is shown. Notice that when $\phi_a = 2n^\prime\pi$, the system is always gapped and a trivial insulator in both cases $\delta t < 0$ and $\delta t > 0$.
\begin{figure}[b]
  \centering
  \includegraphics[width=1\linewidth]{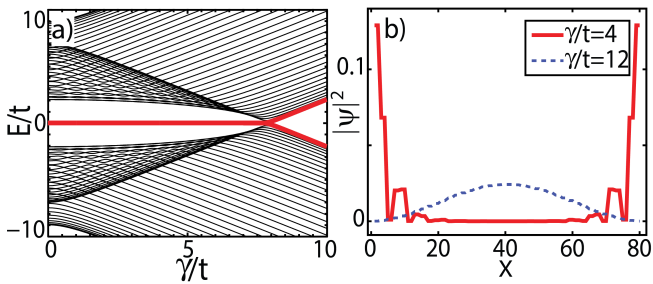}
    \caption{(Color online) (a) The energy spectrum as a function of $\gamma$ under OBCs with 20 unit cells in the presence of staggered perpendicular Zeeman field. (b) Probability distribution for two different values $\gamma/t = 4$ and $12$. Here, $\delta t = -0.5t$, $\theta = \pi/2$, $\lambda = \lambda^\prime = 4t$, and $\phi_a = \pi$.}
    \label{fig6}
\end{figure}

In the mixed regime, for $\delta t < 0(> 0)$, the two-fold degenerate nontrivial phase still remains nontrivial in the presence of staggered perpendicular Zeeman field. The behavior of energy states is depicted in Fig. \ref{fig6}(a) which is similar to the one described in the case of uniform Zeeman field. However, in contrast to the case presented in Fig. \ref{fig4}(a), one can observe that zero-energy states can sustain larger staggered Zeeman field strengths than that of the uniform Zeeman field case. This can be traced back to the specifically staggered magnetic field, having a zero net magnitude in each unit cell, that only deforms the spectrum such that a large value of strength can make a significant effect. Figure \ref{fig6}(b) shows the probability distributions of lowest energy states for different values of $\gamma$ in the presence of staggered perpendicular Zeeman field for the mixed regime.

\section{Winding number}\label{winding}

In addition to the Zak phase, a useful quantity for discussing the bulk features of quantum states is a winding number. It can be used to distinguish different topological phases determining the number of pairs of zero-energy edge states. The definition of the winding number is relevant in the case of block off-diagonal Hamiltonians.
A well-studied feature of systems with the chiral symmetry class, arising from the sublattice or chiral symmetry, is that Hamiltonians describing of such systems can be brought into block off-diagonal form in the basis of chiral operator \cite{Altland,winding}. Obviously, in this basis, the chiral operator $\mathcal{C}$ must be diagonalized, $\hat{U}_1\mathcal{C}\hat{U}^{-1}_1 = -\sigma_z\otimes I$, via the unitary matrix,
\begin{equation}\label{unitary transformation}
\hat{U}_1=\left(
\begin{array}{cccc}
    0&0&1&1\\
     -1&1&0&0\\
     0&0&-1&1\\
     1&1&0&0\\
   \end{array}
   \right).
\end{equation}

It is possible to transform the Hamiltonian (\ref{hk-matrice}) by the unitary transformation $\hat{U}_1$, yielding
\begin{equation}\label{off-diagonal}
\hat{U}_1\hat{h}(k)\hat{U}^{-1}_1=\tilde{h}(k), \quad \tilde{h}(k)=\left(
\begin{array}{cc}
     0&\hat{V}_1\\
     \hat{V}^\dagger_1&0
   \end{array}
   \right),
\end{equation}
with
\begin{equation}\label{off-diagonal V1}
\hat{V}_1(k)=\left(
\begin{array}{cc}
     iM_{y,a}&s(k)+\xi(k)\\
     -\xi(k)^\ast+s(k)^\ast&-iM_{y,b}
   \end{array}
   \right).
\end{equation}

Therefore, the block off-diagonal representation of the Hamiltonian, Eq. (\ref{off-diagonal}), allows for the definition of the chiral index as

\begin{equation}
W = -Tr\int_{-\pi}^{\pi} \frac{dk}{2\pi i}\hat{V}^{-1}_1\partial_k \hat{V}_1=-\int_{-\pi}^{\pi} \frac{dk}{2\pi i}\partial_k Ln Z(k),
\label{winding number}
\end{equation}
with
\begin{align}\label{DetV1}
  Z(k)=&Det\hat{V}_1=-2[(t^2-\delta t^2)+\lambda\lambda^\prime]cos(k)\nonumber \\
  &-2i[(t+\delta t)\lambda+(t-\delta t)\lambda^\prime]sin(k)\nonumber \\
  &-2(t^2+\delta t^2)+M_{y,a}M_{y,b}+\lambda^2+\lambda^{\prime 2}.
\end{align}
$W$ is the winding number of $Z(k)$ characterizing topologically distinct phases. The integration over
$k$ in Eq. (\ref{winding number}) can be performed with the use of Cauchy's residue theorem, leading to a simple analytical formula for the winding number.
For weak and strong modulated spin-orbit coupling regimes as well as mixed regime in the presence of uniform (staggered) perpendicular Zeeman field, we find $W$ as
\begin{eqnarray}
 W = &\Theta&(|M_{y,a}|-\min\{Re\gamma^{u(s)}_{1},Re\gamma^{u(s)}_{2}\})\nonumber\\
 &-&\Theta(|M_{y,a}|-\max\{Re\gamma^{u(s)}_{1},Re\gamma^{u(s)}_{2}\}),
\end{eqnarray}
where $\Theta(x)$ is the Heaviside theta function and $Re$ means the real part. $\gamma^{u(s)}_{1}$ and $\gamma^{u(s)}_{2}$ are defined by Eqs. (\ref{uniZeeman1}) and (\ref{uniZeeman2}) (Eqs. (\ref{stagZeeman1}) and (\ref{stagZeeman2})). $W = 1$ denotes nontrivial topological region which manifest itself by a pair of degenerate zero-energy boundary states under OBCs and $W = 0$ denotes trivial region in which the system is an ordinary insulator. Notice that, the analytical results support the above-obtained numerical topological phases quite well.

\section{Symmetry breaking perturbations}\label{section symmetry}

\begin{figure*}
  \centering
  \includegraphics[width=.7\linewidth]{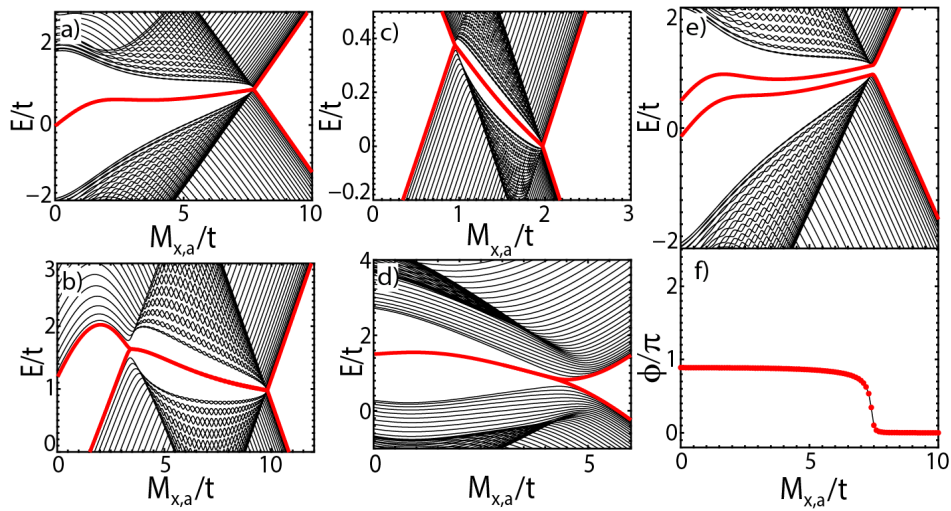}
    \caption{(Color online) Dependence of energy spectrum of a finite lattice with 40 unit cells on $M_{x,a}$ with (a) ((e)) $(M_{x,a} = M_{x,b} - 0.6t)$, $\delta t = -0.5t$, $\lambda = \lambda^\prime = 4t$, $\gamma = 2t$, $\theta = \pi/2$, and $\phi_a = \pi$, (b) $\delta t = -0.5t$, $\lambda = 7t$, $\lambda^\prime = 3t$, $\gamma = 2t$, $\theta = \pi/2$, and $\phi_a = \pi$, (c) $\delta t = -0.5t$, $\lambda = \lambda^\prime = 0.2t$, $\gamma = 0.3t$, $\theta = \pi/2$, and $\phi_a = 0$, (d) $\delta t = 0.5t$, $\lambda = \lambda^\prime = 3t$, $\gamma = 3t$, $\theta = \pi/2$, $\phi_a = \pi$, $V_1 = 2t$, and $V_2 = t$. (f) The corresponding Zak phase of the panel (e).}
    \label{fig10}
\end{figure*}
Now, let us discuss the stability of the topological phase by considering local perturbations which remove the symmetries of the system.
In the nontrivial topological region, in addition to either uniform or staggered $y$-component of Zeeman field, if we introduce the Zeeman field along the $x$-axis with the same direction and strength on the sublattices, $M_{x,a} = M_{x,b}$, then the topological phases fall to the AI class. This is because of breaking of chiral and particle-hole symmetries and preserving of effective time-reversal symmetry. Subsequently, the mid-gap edge states under OBCs are no longer stable and shift away from zero energy value as shown in Fig. \ref{fig10}(a). Moreover, in the absence of $y$-component of Zeeman field, the uniform Zeeman field along the $x$-direction causes the chain to be a nontrivial insulator with Zak phase $\phi_{Zak} = \pi\ (\mathbb{Z} = 1)$ only in the mixed regime under the conditions

 \begin{align}
   \lambda\lambda^\prime > 0&\ \ \ &\textrm{and}& &0< M_{x,a} <|2t|,&\nonumber \\
  \lambda\lambda^\prime < 0&\ \ \ &\textrm{and}& &0< M_{x,a} <|2\delta t|.&
 \end{align}

As discussed in the previous Secs. \ref{section3} and \ref{section4}, the system is an ordinary insulator in the presence of uniform (staggered) Zeeman field along the $y$-axis in the weak (strong) spin-orbit coupling regime within the range $0 < M_{y,a} < \gamma_1^{u(s)}$. Interestingly, in this range, we can close and reopen the energy gap to change the topology of the system with the aid of $x$-component of uniform Zeeman field. The energy gap closure happens in the presence of uniform Zeeman field along the $x$-axis and $y$-component of staggered Zeeman field with conditions
\begin{align}
\gamma^{x}_{1}&=|\lambda-\lambda^\prime|\sqrt{1+\frac{M^2_{y,a}}{4 t^2-(\lambda-\lambda^\prime)^2}},\ \ & \textrm{at}\ \ \ &k = 0,&\nonumber\\
\gamma^{x}_{2}&=|\lambda+\lambda^\prime|\sqrt{1+\frac{M^2_{y,a}}{4\delta t^2-(\lambda+\lambda^\prime)^2}},\ \ & \textrm{at}\ \ \ &k = \pm\pi.&
\label{x1}
\end{align}
Hence, the chain is nontrivial insulator $\mathbb{Z} = 1$ within the range $\gamma^{x}_{1} < M_{x,a} < \gamma^{x}_{2}$. In Fig.  \ref{fig10}(b) the energy spectrum is plotted versus $M_{x,a}$ for $M_{x,a} = M_{x,b}$ and $M_{y,a} = -M_{y,b}$. At $M_{x,a} = 0$, the chain is an ordinary insulator. As we increase $M_{x,a}$, the gap closes at $M_{x,a} = 3.26t$ and low energy states merge together. If the $x$-component of Zeeman field exceeds the certain value $M_{x,a} = 9.79t$, the bulk states join to the mid-gap states and then gap reopens. Similarly, the energy gap closure conditions in the presence of $x$- and $y$-component of uniform Zeeman fields take place at
\begin{align}
\gamma^{\prime x}_{1}&=2|\delta t|\sqrt{1-\frac{M^2_{y,a}}{4\delta t^2-(\lambda+\lambda^\prime)^2}},\ \ & \textrm{at}\ \ \ &k=\pm\pi,&\nonumber\\
\gamma^{\prime x}_{2}&=2|t|\sqrt{1-\frac{M^2_{y,a}}{4t^2-(\lambda-\lambda^\prime)^2}},\ \ & \textrm{at}\ \ \ &k=0.&
\label{x2}
\end{align}
The chain has a nontrivial topology with the condition $\gamma^{\prime x}_{1} < M_{x,a} < \gamma^{\prime x}_{2}$. For this case, the energy spectrum is plotted in Fig. \ref{fig10}(c) which exhibits a nontrivial phase in the parameter space $M_{x,a}\in (0.94t,1.97t)$. Notice that for zero value of the $x$-component of Zeeman field, the relations (\ref{x1}) and (\ref{x2}) reduce to the ones presented in Eqs. (\ref{gap closure k=0}) and (\ref{gap closure k=pi}).

In order to investigate the effect of effective time-reversal symmetry on the stability of the edge states, we add the following perturbation to the total Hamiltonian
\begin{eqnarray}\label{v}
\hat{H}^\prime&=&\sum\limits_{\substack{ n,\\c=(a,b)}}(V_1\hat{c}^\dagger_{n,\uparrow}\hat{c}_{n,\uparrow}
+V_2\hat{c}^\dagger_{n,\downarrow}\hat{c}_{n,\downarrow}),
\end{eqnarray}
where $V_1$ ($V_2$) is the strength of on-site potential. This perturbation breaks not only both particle-hole and chiral symmetries but also effective time-reversal symmetry.
As it is shown in Fig. \ref{fig10}(d) the $x$- and $y$-component of Zeeman fields with perturbation $\hat{H}^\prime$ preserve the mid-gap edge states but shift them away from zero energy. In this case, due to absence of effective time-reversal, chiral and particle-hole symmetries, the topological class is the A.\\

Remarkably, it is easy to obtain that nonuniform Zeeman field along the $x$-axis, $M_{x,a} \neq M_{x,b}$, breaks the inversion symmetry even in the presence of either uniform or staggered perpendicular Zeeman field. In Fig. \ref{fig10}(e)((f)), the energy spectrum (Zak phase) is depicted versus $M_{x,a}$ for different magnitudes of $x$-component of Zeeman fields on each sublattice, {\it i.e.}, $(M_{x,a} = M_{x,b} - 0.6 t)$. In this case, the edge states separate from each other and, subsequently, the Zak phase does not have quantized values. Therefore, the above inspections insure that the topological edge states in the studied system, belonging to the BDI, AI, and A classes, are protected by the inversion symmetry.

%%%%%%%%%%%%%%%%%%%%%%%%%%%%%%%%%%%%%%%%%
\section {Conclusions} \label{section5}
%%%%%%%%%%%%%%%%%%%%%%%%%%%%%%%%%%%%%%%%%

We have investigated the mutual effects of Zeeman field and modulated spin-orbit coupling on the topological characteristics of 1D dimerized lattice. It is shown that 1D lattice dimerization is a trivial insulator in different parameter regions of weak and strong modulated spin-orbit coupling. We have also showed that with the aid of uniform perpendicular Zeeman field for weak modulated spin-orbit coupling and staggered perpendicular Zeeman field for strong modulated spin-orbit coupling, the trivial phases change to a nontrivial topological phase with a topological phase transition. Therefore, as compared with the bare SSH model, the region of nontrivial topological phase in the parameter space can be extended by adding the modulated spin-orbit coupling and the perpendicular Zeeman field to the SSH model. This nontrivial topological phase exhibits one pair of zero-energy edge states obeying non-Abelian statistics that is a key requirement in topological quantum computing. The topological phases of the system are classified with $\mathbbm{Z}$ number and belongs to the BDI class. We also have derived an analytical formula for the winding number through bulk quantum states to confirm our numerical results. Finally, the stability of the edge states to externally applied Zeeman field along the chain and on-site spin-dependent potential has also been investigated. The inclusion of these effects breaks chiral, particle-hole, effective time-reversal, and inversion symmetries resulting in that the edge states are topologically protected and robust against the perturbations as long as the inversion symmetry requirement is satisfied.

We remark that the discussion presented here is an example of the general issue implying that required experimentally accessible regions in parameter space may be engineered, in order to reach desired phases efficiently.

\section*{Acknowledgment}
The authors would like to thank H. Guo, Y.-M. Lu, and Zh. Yan for valuable comments. We are also grateful to A. Najafi for reading the manuscript carefully. This work is partially supported by Iran Science Elites Federation under Grant No. 11/66332.%The authors acknowledge financial support from Iran Science Elites Federation under Grant No. 11/66332.

\end{document}